\documentclass[useAMS,usenatbib]{biom}
\usepackage{amssymb}
\usepackage{amsfonts}
\usepackage{amsmath}
\usepackage{appendix}
\usepackage{graphicx}
\usepackage{caption}
\usepackage{bm}
\usepackage{bbm}
\usepackage{color}
\usepackage{xr}
\def\bSig\mathbf{\Sigma}

\newcommand{\eps}{\epsilon}
\newcommand{\boldZ}{{\boldsymbol Z}}

\newcommand{\IND}{{\mathbbm{1}}}

\title[Effect Sizes of cis-eQTLs Using a Log of Linear Model]{Estimation of cis-eQTL Effect Sizes Using a Log of Linear Model}

\author{John Palowitch$^{1,*}$\email{palojj@email.unc.edu}, 
Andrey Shabalin$^{2}$,
Yi-Hui Zhou$^{3}$,\\
{\bf  Andrew B.\ Nobel$^{1,**}$\email{nobel@email.unc.edu}, and 
Fred A.\ Wright$^{4,***}$\email{fred\_wright@ncsu.edu}}\\
$^{1}$Department of Statistics and Operations Research, University of North Carolina at Chapel Hill,\\
Chapel Hill, North Carolina, U.S.A.\\
$^{2}$Center for Biomarker Research and Personalized Medicine, Virginia Commonwealth University,\\
Richmond, Virginia, U.S.A.\\
$^{3}$Bioinformatics Research Center and Department of Biological Sciences, North Carolina State University,\\
Raleigh, North Carolina, U.S.A.\\
$^{4}$Department of Statistics, North Carolina State University}

\begin{document}





\pagerange{\pageref{firstpage}--\pageref{lastpage}} 
\volume{--}
\pubyear{2016}
\artmonth{--}




\label{firstpage}


\begin{abstract}
The study of expression Quantitative Trait Loci (eQTL) is an important problem in genomics and biomedicine.
While detection (testing) of eQTL associations has been widely studied, less work has been devoted to the estimation of
eQTL effect size. 
To reduce false positives, detection methods frequently rely on linear modeling of rank-based normalized or log-transformed gene expression data.
Unfortunately, these approaches do not correspond to the simplest model of eQTL action, and thus yield estimates of eQTL association that can be uninterpretable and inaccurate.
In this paper we propose a new, log-\emph{of}-linear model for eQTL action, termed ACME, that captures allelic contributions 
to cis-acting eQTLs in an additive fashion, yielding effect size estimates that correspond to a biologically coherent model of cis-eQTLs.   
We describe a non-linear least-squares algorithm to fit the model by maximum likelihood, and obtain corresponding $p$-values. 
We perform careful investigation of the model using a combination of simulated data and data from the Genotype Tissue Expression (GTEx) project. Our results reveal little evidence for dominance effects, a parsimonious result that accords with a simple biological model for allele-specific expression and supports use of the ACME model. We show that Type-I error is well-controlled under our approach in a realistic setting, so that rank-based normalizations are unnecessary. Furthermore, we show that such normalizations can be detrimental to power and estimation accuracy under the proposed model.
We then provide summaries of ACME effect sizes for whole-genome cis-eQTLs in the GTEx data. 

\end{abstract}

%



\maketitle


%

\section{Introduction}\label{Introduction}
An expression Quantitative Trait Locus (eQTL) is a genetic polymorphism (typically a single-nucleotide 
polymorphism, abbreviated by SNP) that is associated with transcriptional expression levels
in a particular tissue.  The statistical analysis of eQTLs has become increasingly important in 
understanding molecular mechanisms connecting genetic variation to complex traits and disease (\cite{morley2004genetic, gilad2008revealing, grundberg2012mapping, westra2013systematic}). For example, eQTL studies can be used to plausibly link disease phenotypes analyzed in Genome-Wide Association Studies (GWAS) to gene expression, with recent work focusing on variation across tissues 
\citep{gamazon2015gene, ardlie2015genotype}. 
Although the underlying biology is complex, a fundamental step in many analyses is to compare the genotypes of a large number of SNPs to the expression levels of all genes, which presents challenges in computation and multiple testing \citep{wright2012computational}.

Statistical analyses of eQTLs have often been based on standard linear regression 
\citep{shabalin2012matrix}, 
with a focus on testing and detection. 
A key step, commonly considered necessary to avoid false positives, has been to normalize and transform the 
expression data prior to analysis \citep{beasley2009rank}. Often normalization removes the scale of the expression data, and with it, a natural measure of effect size due to genotype.
As a consequence, eQTL effect size has often been described in terms of regression partial $R^2$ between genotype and transformed expression (see for example \cite{stranger2007population}). 
However, the $R^2$ statistic can be highly sensitive to transformations of the response, and is difficult to interpret biologically. An appropriate eQTL model should reflect a coherent model of allelic contributions to expression, and provide a null hypothesis to test for dominance \citep{powell2013congruence}. Furthermore, a biologically appropriate effect-size model will improve the accuracy of hypothesis tests (as we show in this paper), and provide reliable rankings of eQTLs in terms of effect sizes instead of $p$-values alone.

In this paper we propose ACME, a model for the effect size of cis-acting eQTLs, in which the effects of genotype alleles on expression 
are \underline{A}dditive \underline{C}ontributions on the original expression scale, with 
\underline{M}ultiplicative \underline{E}rror.  
In the ACME model the log of expression is equal to the {\em log of a linear systematic term} 
(``log-of-linear") plus noise and covariate effects: this seemingly subtle difference from
standard log-linear modeling is of key importance in estimating and interpreting effect sizes. 
The ACME model reflects a marked departure from standard practice 
in eQTL analyses and has important implications for downstream inferences on effect sizes and  dominance.
Standard normalizations meant to control false positives run against, as we will argue, the most coherent conception of eQTL action: namely that gene expression is \emph{additive} in allele count. A primary contribution of this paper is to assess the validity of this conception against the alternatives suggested by standard practices. Additionally, we provide (i) a fast, custom fitting algorithm and corresponding software package, (ii) a robustness analysis of our method, and (iii) diverse results and comparisons from a full cis-eQTL analysis using both ACME and existing models.
	


The organization of the paper is as follows. In the remainder of this section, we discuss 
 standard eQTL models. In Section \ref{Model}, we lay out the ACME 
model and conduct statistical tests on real data to show its conformity to cis-eQTL 
action. In Section \ref{pvalues} we analyze robustness of ACME p-values to violations of model assumptions in real data. Section \ref{Power} describes a simulation study to assess  consequences of using standard normalizations when ACME is the true model. In Section \ref{RealData} we discuss results from analyses of all cis-eQTLs in nine tissues (data from the GTEx project, \citealt{lonsdale2013genotype}) using both ACME and existing methods. In Section \ref{Discussion}, we summarize our contributions and discuss future research.




\subsection{Existing approaches to gene expression modeling}\label{existing}

Gene expression data are rarely analyzed on the original scale, due to heteroskedasticity and heavy-tailed errors \citep{rantalainen2015robust}. Instead, logarithmic transformation of expression is a standard pre-processing step for many microarray platforms 
\citep[e.g.][]{morley2004genetic, irizarry2003exploration} and often plays an important role in
downstream analyses such as differential expression  
\citep[e.g.][]{cancer2013genomic, li2014regression}.
RNA-Seq data are inherently count-based, and statistical analyses of such data often make use of 
binomial, negative-binomial, or Poisson generalized linear models \citep{mccarthy2012differential, zhou2011powerful, zwiener2014transforming}, which use logarithmic or near-logarithmic link functions.  
However, count-based modeling is rarely used in eQTL analysis, due to computational requirements, and the fact that several stages of read-count normalization are usually applied to expression data. Furthermore, count-based modeling may not be necessary in studies with large sample sizes \citep{zhou2011powerful}, and eQTL analyses are often performed using linear 
regression of log-transformed expression, assuming additive allelic effects on the log scale \citep[e.g.][]{myers2007survey}.

Another common transformation of expression is inverse quantile-normalization \ \citep[e.g.][]{dixon2007genome}, used to ensure normality of residuals under the null \citep{beasley2009rank, szymczak2013adaptive}. Given a vector $y$ of length $n$, the quantile-normalization (QN) transformation is the function $Q(y_i)=\Phi^{-1}((rank(y_i)/(n+1))$, mapping each value to a normal quantile corresponding to its rank. Henceforth, eQTL analysis involving linear regression of quantile-normalized gene expression will be referred to as ``QN-linear". For eQTL analyses, the QN-linear model yields $p$-values that are approximately uniform under the null of no association between 
expression and genotype.  However, the quantile-normalization mapping inherent to this approach erases all connection between the linear model parameters and the original gene expression. Hence,  estimated coefficients from QN-linear regression do not reflect the scale of the original 
data, and contain almost no information about the true allelic effect.
As a result, QN-linear model effect-size estimates from eQTLs with clearly diverse signal-to-noise ratios can yield nearly identical $p$-values (see Web Appendix \ref{qn-vs-ACME}).

\subsection{Notation and data}\label{notation}
eQTL data from $n$ samples will be written as follows.
The SNP genotype is the number of minor alleles 0, 1, or 2, rounded if using imputed data. 
Genotype is contained in an $S \times n$ matrix where $S > 0$ is the number of SNP markers; we denote a
(length $n$) row of the genotype matrix by $s$.  
Expression is measured by the number of mapped reads relative to the 
library size (see Web Appendix \ref{data_prep}).
Read counts for expression are contained in a $T \times n$ matrix 
where $T>0$ is the number of genes or transcripts; the
length $n$ row of expression matrix is denoted $c$. 
Finally, the $p$ covariates (e.g. sex or batch) are stored in a $p \times n$ matrix. 

The GTEx pilot data set \citep{ardlie2015genotype} is used for all analyses and investigations. Given the purported scope of the ACME model, the analysis focus is on ``cis" gene-SNP pairs, for which the SNPs are within 1 megabase upstream or downstream of the transcription start or stop sites. The GTEx pilot data contains a SNP database and expression data from nine tissues, each having sample sizes between $n = 83$ and $n = 156$. Each tissue-specific data set had $p = 19$ covariates: sex and 3 genotype principal components, which were shared across all tissues; and 15 PEER (Probabalistic Estimation of Expression Residuals) factors computed from expression data \citep{stegle2012using}.




\section{The ACME model and diagnostics}\label{Model}
This section provides a ground-up introduction of the ACME model. The first consideration is the appropriate scale of expression data for error control. As discussed in Section \ref{existing}, errors from linear models of raw gene expression data are known to be 
heteroskedastic and non-normal. In Web Appendix \ref{raw-expr-resids},
we show that the non-normality observed in real-data residuals after linear regression with raw expression causes severe Type-I error discrepancies. Though the QN-linear model avoids this problem, it is unsuitable for effect-size estimation (as discussed in Section \ref{existing}). Thus we assess the commonly-used log transformation, in two ways. First, we display tests of normality and heteroskedasticity of residuals after 
fitting QN-linear, standard linear, and various log-linear models (see Web Appendix \ref{normality}) 
to GTEx data. 
We see that models using log-transformed expression perform much better than those based on 
raw expression.  Comparison to the QN-linear model results indicate that the log-transformation is
still somewhat subject to noise and outliers.  However, we consider the resulting violations of normality and homoskedasticity 
to be acceptably modest, when balanced against the ability to assess effect size with a model that respects the scale of expression data. (Section \ref{pvalues} provides a deeper look into Type-I errors from log-based methods.) Second, we assess residual normality under the box-cox transformation \citep{box1964analysis}, defined for $\lambda\in\mathbb{R}$ as
\begin{equation}\label{boxcox}
t_\lambda(y) = \begin{cases}\tfrac{y^\lambda - 1}{\lambda},&\lambda\ne0\\\log(y),&\lambda=0.\end{cases}
\end{equation}
We perform Shapiro-Wilk tests on box-cox transformed expression data from null eQTLs (as judged by QN-linear p-values) subsampled from real data. We find that the log transformation ($\lambda = 0$) consistently results in the fewest instances of significantly non-normal residuals. These results are displayed in Web Figure \ref{fig:box-cox}.

The analyses described above suggest that the log transformation puts gene expression on a natural scale for error control in eQTL effect-size analysis. We now discuss systematic components of various log-scale effect size models, including ACME (to be introduced). Henceforth, let $y_i := \log(1 + c_i)$ denote the log-transformed normalized gene read count from sample $1 \leq i \leq n$, where the addition of 1 avoids taking the logarithm of zero. 
The value $c_i$ is the result of taking the original raw count for the gene in sample $i$, library-normalizing and then scaling up
to the magnitude of the original mean count (see Web Appendix \ref{data_prep}).
Let $s_i$ denote the minor allele count for the SNP in sample $i$ ($s_i \in \{0, 1, 2\}$).
Let $\boldZ_i$ denote the $p \times 1$ vector of covariates for sample $i$, and let $\gamma$ be 
an unknown $p \times 1$ covariate coefficient vector.  Finally, let $\eps_1, \ldots, \eps_n$ be independent 
$N(0, \sigma^2)$ errors with positive variance $\sigma^2$. 
Note that the quantities $\sigma$ and $\gamma$ may differ across gene-SNP pairs. 
It is common in eQTL studies to assume that the covariate effect ${\mathbf Z}_i^T \gamma$ contributes to 
expression on the same scale as the noise \citep[e.g.][]{shabalin2012matrix}.  We follow this practice for all models considered in this paper (including linear regressions with both raw and quantile-normalized gene expression). Additional support for this convention can be seen from the fact that the covariates were computed from normalized data to reduce the influence of outliers \citep{ardlie2015genotype}, so it is natural for them to be residualized on the log-scale.

\subsection{Log-ANCOVA and log-linear models}\label{log-models}
We now describe two log-scale linear eQTL models. If one assumes each genotype is associated with a 
distinct level of average log-expression, the associated linear model effectively includes a dominance term for 
the homozygous genotype for the reference allele, 
and yields what we call
the ``log-ANCOVA" model:
\begin{equation}\label{eq:logANCOVA}
y_i = \alpha_0\IND_0(s_i) + \alpha_1\IND_1(s_i) + \alpha_2\IND_2(s_i) + \boldZ_i^T\gamma + \eps_i .
\end{equation}
Here the parameters $\alpha_j$ are unknown log-expression means corresponding to the genotypes, 
and $\IND_k(s_i) = 1$ if $s_i = k$, and zero otherwise. Another (simpler) log-scale model includes just one parameter for allele \emph{count}:
\begin{equation}\label{eq:logLinear}
y_i = \theta_0 + \theta_1s_i + \boldZ_i^T\gamma + \eps_i.
\end{equation}
Above, $\theta_0$ is baseline log-expression, and $\theta_1$ is the contribution to log-expression of each reference allele. This model includes one fewer degree of freedom than log-ANCOVA, due to the loss of the dominance term $\alpha_2$. Linear regression on allele count 
has been heavily used in eQTL analysis \citep{ardlie2015genotype},
perhaps partly because evidence of eQTL dominance effects are scant, even in trans-analyses \citep{wright2014heritability}. Furthermore, simpler models like \eqref{eq:logLinear} are useful in that they can be used to test for dominance and, in case of allelic independence, provide appropriate estimates of eQTL action. 



\subsection{The ACME Model}\label{ACME}
Despite the prevalence of the log-linear model \eqref{eq:logLinear} in eQTL analysis, it has not been subjected to careful scrutiny. The current understanding of cis-eQTL variation in humans is that it is largely allele-specific 
\citep{castel2015tools}, i.e.,
the transcription of a gene in a particular chromosome is influenced primarily by one or more 
SNP alleles on the same chromosome.
Thus, in the absence of feedback mechanisms, the effect of each SNP allele should
be additive on the {\it original} expression scale.
To incorporate this understanding while respecting the heavy-tailed nature of gene expression data,
we propose the following log-scale non-linear regression
(ACME):
\begin{equation}\label{eq:ACME}
y_i = \log(\beta_0 + \beta_1s_i) + \boldZ_i^T\gamma + \eps_i.
\end{equation}
Here $\beta_0$ is the baseline mean expression, and  $\beta_1$ the additive contribution of 
each allele. 
Exponentiating each side of equation \ref{eq:ACME}, on the expression
scale we have:
\begin{equation}\label{eq:ACME-exp}
c_i + 1 = \big(\beta_0 + \beta_1s_i\big) \;\cdot\; \exp\left(\boldZ_i^T\gamma + \eps_i\right).
\end{equation}
It is clear from this equation that the effect of genotype is linear in raw expression, as desired{\color{red}.}
We fit the ACME model to data via maximum likelihood, using a Gauss-Newton algorithm, which is derived 
in Web Appendix \ref{Fitting}.

\subsubsection{The effect size}
The coefficients $\beta_1$ and $\beta_0$ from the ACME model operate on the original expression scale, 
so they lend themselves naturally to a ``fold-change" interpretation. 
In particular, the ratio
$\beta_1/\beta_0$ represents the fraction of mean increase due to a single referent allele compared to
the baseline genotype 0.  We note that Equation \eqref{eq:ACME} may be written 
\begin{equation}\label{eq:ACME2}
y_i = \log(\beta_0) + \log\left(1 + \frac{\beta_1}{\beta_0} s_i\right) + \boldZ_i^T\gamma + \eps_i,
\end{equation}
which separates the role of $\beta_0$ in determining baseline expression and 
the role of $\beta_1/\beta_0$ in determining the effect of genotype. 
Equation \ref{eq:ACME2} also plays a role in the fitting algorithm. 
In what follows we use ``effect size" to refer to the ratio $\beta_1/\beta_0$. 
In Web Appendix \ref{SE} 
we obtain a formula for the standard error of $\beta_1/\beta_0$, using a reduced 
Hessian matrix derived from the model. 
We note that other notions of effect size may be of interest to biologists. For example, an alternative effect size model, studied simultaneously and independently, incorporates allele additivity through an explicit focus on a fold-change parameter \citep{mohammadi2016quantifying}.

\subsubsection{Fitting algorithm and software}
Though the ACME model can in principle be fit by brute-force likelihood maximization, we found stock implementations of this approach to be slow and unreliable in practice. We have crafted a custom fitting algorithm for ACME, provided in Web Appendix E.
We have implemented the fitting algorithm, and a parallelized wrapper for full cis-genome analysis, in a free and open-source software package called $\mathtt{ACMEeqtl}$. The package is written in the $\mathtt{R}$ statistical computing language, and is available on the CRAN repository. In Section \ref{Power:computation}, we benchmark the computation time of our software.




\subsection{Goodness-of-fit tests}\label{diagnostics}
In the previous section, we pointed out that
the log-linear model assumes allelic effect additivity on the log-expression scale, whereas ACME assumes additivity on the raw-expression scale. These assumptions can be treated as competing hypotheses, evaluations of which may be performed with goodness-of-fit tests. In this section, we carry out such tests using data from the GTEx project.
To derive the test, note that the log-linear and ACME models are each nested within the log-ANCOVA model. For instance, the log-ANCOVA model reduces to the ACME model via the parameterization
\begin{align*}
&\alpha_0(\beta) := \log(\beta_0),\\
&\alpha_1(\beta) := \log(\beta_0) + \log\left(1 + \frac{\beta_1}{\beta_0}\right),\\
&\alpha_2(\beta) := \log(\beta_0) + \log\left(1 + 2\frac{\beta_1}{\beta_0}\right).
\end{align*}
Thus, if either 
model is sufficient to explain variation in gene expression, any further improvements in the log-ANCOVA fit should be small and consistent with the extra degree of freedom in that model. Conversely, if the fit of log-ANCOVA is (significantly) better than a smaller model, it suggests the smaller model is insufficient. In testing sets of coefficients in nonlinear regression models with normal errors, $F$-tests are widely used \citep{smyth2002nonlinear}, and generally better handle the degree of freedom issues posed by numerous covariates than do likelihood ratio tests.
Define $SSE_3$ as the sum of squared residuals from the fit of log-ANCOVA, and $SSE_2$ as the sum of squared residuals from the fit of any nested model with 2 degrees of freedom (e.g.\ LL and ACME). Then the goodness-of-fit test statistic is $F = \frac{SSE_2 - SSE_3}{SSE_3 / (n - p - 3)}$ which is approximately $F$-distributed with 
$1$ and $n - p - 3$ degrees of freedom. 
A $p$-value for the goodness-of-fit test is then obtained from the upper-tail of $F_{1,n-p-3}$.

\begin{figure}[!htb]
\centering
\makebox{
	\includegraphics[scale = 0.16]{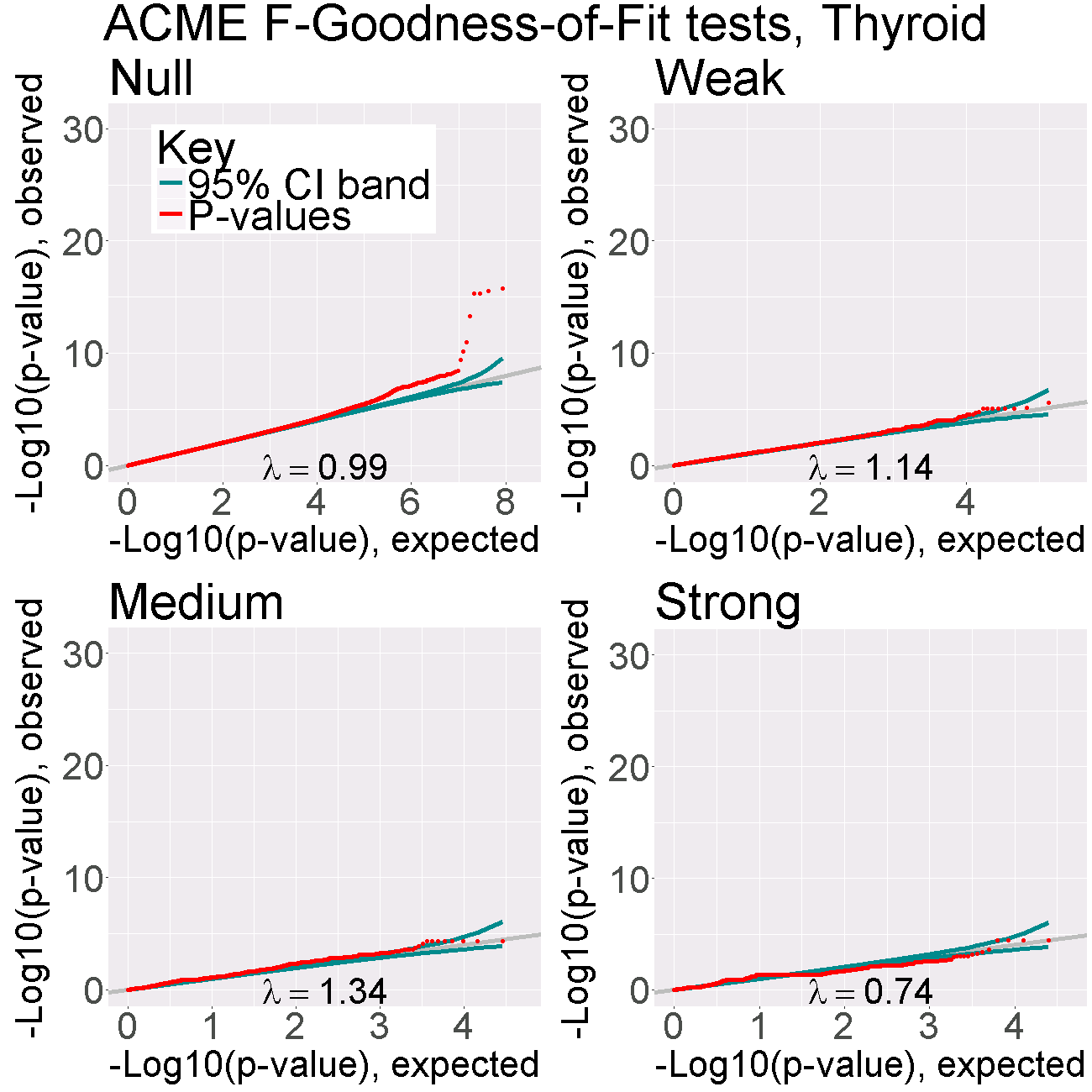}
}
\vskip0.2cm
\makebox{
	\includegraphics[scale = 0.16]{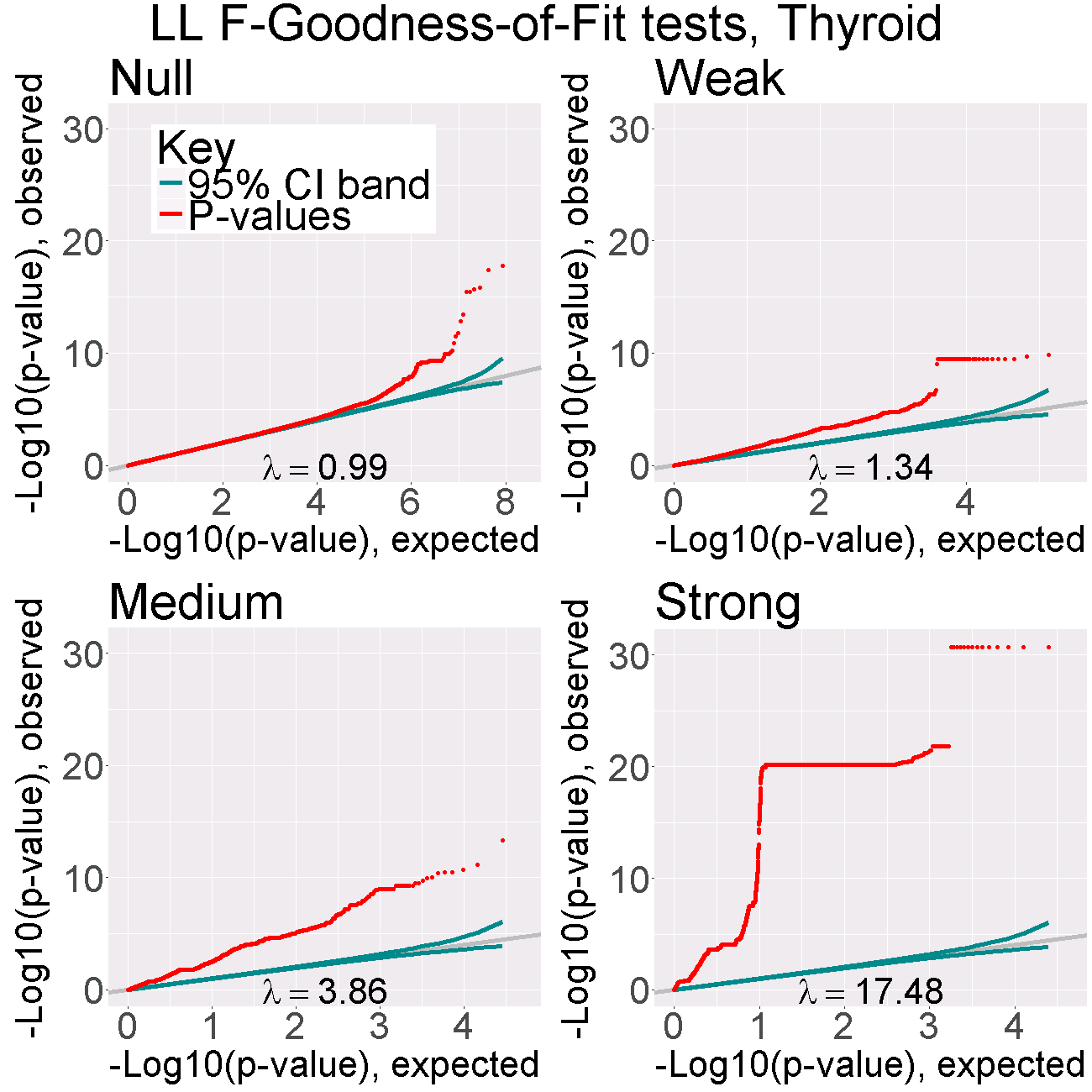}
}
\vskip0.2cm
\makebox{
	\includegraphics[scale = 0.16]{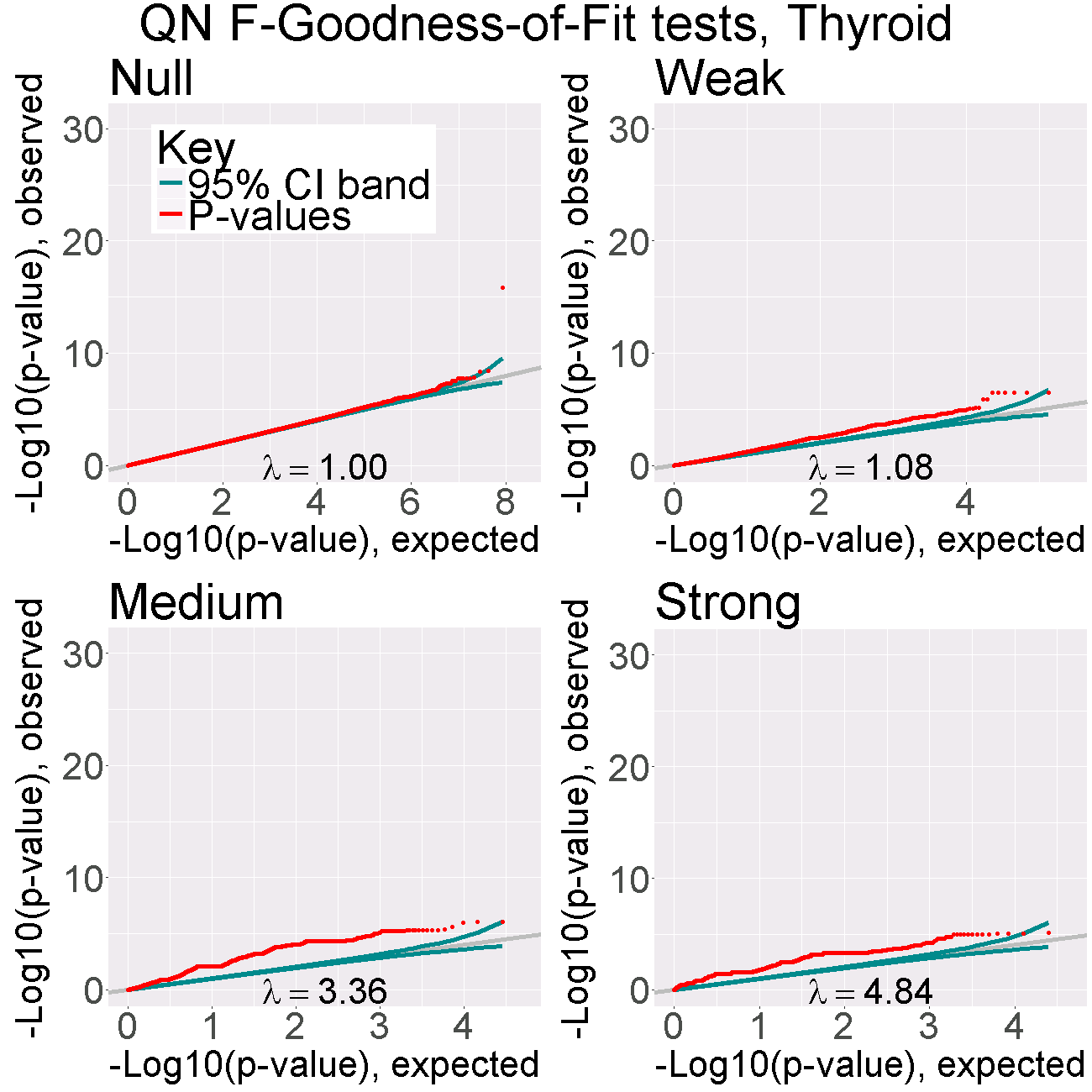}
}
\caption{\label{fig:diagnostic2} Q-Q plots of likelihood ratio test $p$-values for ACME, LL, and QN models, in each sector of GTEx Thyroid sample data, $n = 105$. The grey line is where we would expect the $p$-values (represented by the red dots) to fall if they were perfectly uniform, and the green line represents the 95\% window of error around this expectation. $\lambda$ is the estimated genomic inflation factor.}
\end{figure}

For both the log-linear and ACME models, we applied the goodness-of-fit $F$ test to every cis-acting gene-SNP
pair in Thyroid tissue (with $n = 105$ tissue samples) from GTEx pilot data. As the log-linear and ACME models are indistinguishable under the null model ($\beta_1=0$), we examined the distribution of goodness-of-fit p-values on four bins of cis-eQTL strength (``Null", ``Weak", ``Medium", and ``Strong"), as judged by QN-linear model regression p-values (described fully in Web Appendix \ref{sampling-scheme}). To judge the distribution of the $F$-test p-values from each model, we plotted Q-Q plots on the $-\log_{10}$ scale (see Figure \ref{fig:diagnostic2}). On each plot, we also supplied the genomic inflation factor $\lambda$. The genomic inflation factor is defined by $\lambda:= \text{median}_i\{\chi^2_i\} / 0.455$, 
where $\chi^2_i$ is the 1 d.f. chi-squared test statistic corresponding to the $p$-value for the $i$-th test, following the original reasoning for genomic control \citep{devlin1999genomic}.
Figure \ref{fig:diagnostic2} shows that the distribution of $F$-statistic $p$-values from the log-linear 
model grow increasingly non-uniform as eQTLs become more significant, suggesting that the log-linear 
model is mis-specified.
In contrast, $F$-statistic $p$-values for the ACME model are approximately uniform 
for eQTLs of all strengths.
In other words, the fit of the ACME model is largely indistinguishable from that of log-ANCOVA, whereas the fit of the log-linear model is largely insufficient to explain non-null eQTLs. Overall, these results provide strong empirical support for the raw-expression allelic additivity assumption of the ACME model, and against the log-expression additivity of the log-linear model. We conclude that, among models nested within log-ANCOVA (which are all models based solely on allelic effect), ACME best conforms to the underlying eQTL signal. Similar results were observed after applying the same sub-sampling and testing pipeline to data from four other GTEx pilot tissues (see Web Figures \ref{fig:more-gof-acme} and \ref{fig:more-gof-ll}). For completeness, we also applied the above goodness-of-fit pipeline to the QN-linear model, using the corresponding ANCOVA with quantile-normalized expression. We find the same upward trend of poor fits with stronger effect size observed in the LL model. Along with the discussion in Section \ref{existing}, this further illustrates the inadequacy of quantile-normalized linear models to capture eQTL action.

\section{Model $p$-values and Type I error}\label{pvalues}

In this section we address violations of residual normality, which can affect Type I error. The large number of tests performed in eQTL analyses presents a special challenge for false positive control. For cis-analysis, the number of tests is typically on the order of $10^7$ \citep{lonsdale2013genotype}. Thus, using a Bonferroni bound to control family-wise error at 0.05 requires $p$-values
to be accurate at values of $10^{-9}$. In order to perform a test of no effect, i.e., 
$H_0:\frac{\beta_1}{\beta_0} = 0\;\;\;\text{vs.}\;\;\;H_a:\frac{\beta_1}{\beta_0}\ne 0$
for a given gene-SNP pair, we fit the ACME model and the reduced mean-model with $\beta_1 = 0$, and then
derive a p-value by comparing the resulting $F$-statistic with the $F_{1, n - p - 2}$ distribution.
Non-normality in errors can potentially result in non-uniform $F$-statistic $p$-values under the null. 
To assess this, we examined the performance of ACME on simulated null data with realistic errors. 
In addition, we examined the effect of skew in errors for the extremal $p$-values resulting from large numbers of tests.




\subsection{Empirical performance of the $F$ test}\label{F-test-sims}

\begin{figure}[!htb]
\centering
\mbox{
\includegraphics[scale = 0.10]{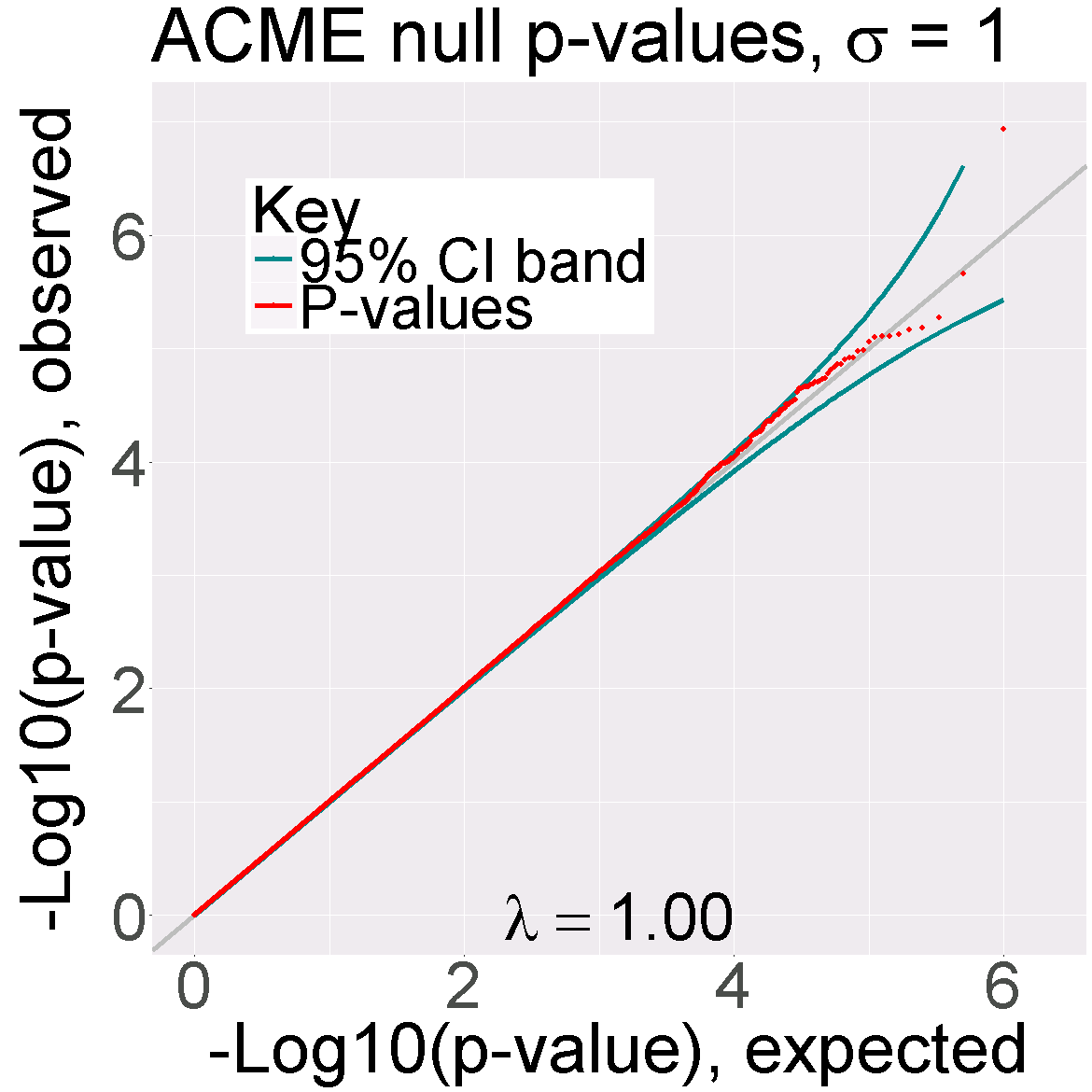}
\includegraphics[scale = 0.10]{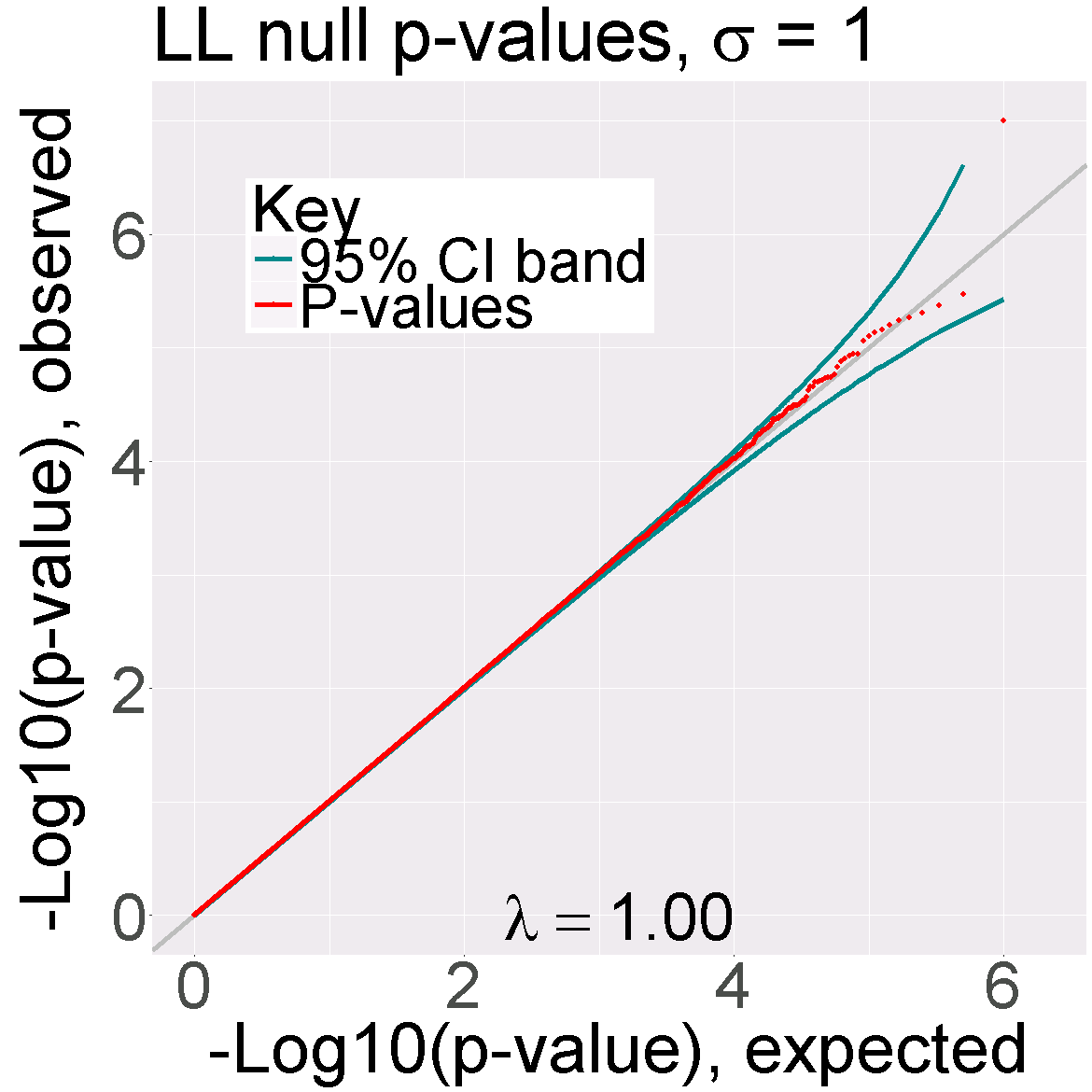}
}
\caption{\label{fig:null_pvals} $p$-value distributions from null simulated data with realistic errors and real covariate/genotype data. $\lambda$ values are inflation factors.}
\end{figure}

We began our investigation of the empirical performance of the $F$ test by fitting the ACME model to null 
data simulated with realistic residuals.  The residuals for each simulated gene-SNP pair were obtained by re-sampling estimated residuals from ACME fits to real GTEx data (full details in Web Appendix \ref{sec5app}). 
Both the ACME and log-linear models were fit to 1 million null eQTLs generated in this manner, 
and $p$-values were obtained from each method using the $F$-test. 
The results are shown in Figure \ref{fig:null_pvals}.  The $F$-test $p$-values appear
nearly uniform for both models, as the inflation factors were in the range 0.995-1.005 ($\lambda = 1$ corresponds to no inflation). 
We emphasize that these conclusions address the behavior of the ACME fit under a realistic null $-$ the earlier analyses established that ACME offers superior fit for real data when evidence of the alternative is strong. Similar results for different settings of the variance of the simulated error are shown in Web Figure \ref{fig:null_pvals-extra}.

\subsection{Importance sampling estimates of Type I error under skew in residuals}\label{importance-sampling}

While the results above are encouraging, consideration of 1 million null pairs is not sufficient to assess 
the quality of very small $p$-values under realistic errors. 
We are unaware of any attempts via direct data simulation to quantify robustness in eQTL studies to 
the stringent multiple testing thresholds necessary for eQTL studies (as low as $10^{-9}$ for cis-testing). 
We note that the robustness investigations of \cite{rantalainen2015robust} used only $10^6$ simulations 
for each investigated condition.

A computationally efficient way to assess Type-I error rates for extreme nominal $p$-values is to perform 
importance sampling \citep{tokdar2010importance}, in which samples are drawn from an appropriate 
alternative distribution (with $\beta_1 \ne 0$), then re-weighted to provide an estimated probability
of rejection under the null. 
For regression models, skewness in the error distribution 
has a major impact on false positive control, and can cause both conservative or
anti-conservative behavior \citep{zhou2015hypothesis}.
Accordingly, we carried out importance sampling using the skew-normal model \citep{azzalini1996multivariate} for the 
distribution of $\epsilon$, with skewness determined by a parameter $\delta$, with $\delta=0$ corresponding 
to the assumed normal error model.  Simulations were performed with modest average expression 
$\beta_0=100$, error variance $\sigma^2=1$, minor allele frequencies 0.025, 0.05, and 0.1, and for sample sizes 
$n=$ 100, 250, and 500.  Among the GTEx datasets used in this paper, most showed skewness in ACME 
residuals between -0.5 and 0.5 (see Web Figure \ref{fig:skew-results} for an example using Adipose GTEx pilot data). 
We chose the skew-normal parameter to correspond to skewness in this range (details in Web Appendix \ref{sec5app2}). 
Target
type I error values $\alpha$ ranged from $10^{-20}$ to $10^{-1}$.

The results for the ACME $F$-test $p$-values are shown in Web Figures \ref{fig:investigate-skew1}-\ref{fig:investigate-skew3}, using the importance sampling 
approach detailed in Web Appendix \ref{sec5app2}.
Some general conclusions can be drawn. For $n=100$ and 
negative skewness in $\epsilon$ (with $\delta = -0.45$), the ACME $p$-values are noticeably conservative 
for $\alpha< 10^{-6}$. For positive skewness, the $p$-values are slightly anti-conservative, but
more accurate than for negative skewness due to asymmetry in the behavior of the systematic 
component of the ACME model. For larger $n=250$, the conservativeness under negative skewness is less extreme, and the $p$-values reasonably accurate to $\alpha=10^{-9}$ for the skewness range shown. This suggests that p-values for the ACME
model should produce acceptable Type-I error rates for most gene-SNP pairs, even for those with relatively small sample sizes. Even pairs with higher skew show acceptable type I error for sample sizes of 250 or greater, and any
deviations from ideal behavior tend to be conservative. These trends hold across the tested values of the MAF, though the p-value skew becomes more severe for lower MAF. For trans-analysis, larger sample sizes may be required due to the more stringent testing thresholds. However, \cite{wright2014heritability} suggested that sample sizes $>1000$ are necessary to reliably detect trans-eQTLs, and for such large studies we would expect robust ACME $p$-values, a separate issue from whether ACME is appropriate for trans-analysis. For small sample sizes and to serve as an ancillary approach, we describe the MCC method \citep{zhou2015hypothesis} as a fast method to obtain robust $p$-values in Web Appendix \ref{mccapp}.  




\section{Power, estimation accuracy, and computation speed}\label{Power}
In this section we present simulation results which display the detection power, estimation accuracy, and computation speed of the ACME model versus existing alternatives. Recall the representation of the ACME model from equation \ref{eq:ACME2}, involving the parameter $\eta:=\beta_1/\beta_0$. We simulated 100 repetitions of the model at values of $\eta$ along the range $(-0.5, 10)$.
The sample-size was set to $n = 105$, as components of the simulations were taken from real data (as in Section \ref{F-test-sims}). At each repetition, the other components of the model were set as follows: (1)
allele counts from a randomly sampled real-data allele count vector corresponding to GTEx samples of Thyroid tissue; (2) real-data covariate matrix corresponding to Thyroid samples, constant across all repetitions and values of $\eta$; (3) noise vector ($\eps_{n\times 1}$) and covariate effect ($\gamma_{p\times 1}$) generated as normals with mean 0 and covariances $\sigma_\eps^2 I_n$ and $\sigma_\gamma^2 I_p$ (respectively), independent within and across repetitions.

We replicated the above simulation framework for various choices of $\sigma_\gamma$, with $\sigma_\eps$ fixed at 1. We then applied linear (RAW), quantile-normalized linear (QN), log-linear (LL), log-ANCOVA (ANCOVA), and ACME models to each instance of the simulation. For each value of $\eta$, we computed the average and standard deviation over the repetitions of the following metrics (per model). (1) $F$-test p-value for hypotheses $H_0:\eta = 0$ vs.\ $H_1:\eta\ne 0$; (2) Estimated raw expression value when reference allele count equals 1; (3) Estimated raw expression value when reference allele count equals 2.
Note that estimation with the QN model cannot provide (2) or (3), as the model is based on a rank-normalized expression. Furthermore, parameters of the LL model are not directly comparable to those of ACME or RAW (as they are additive on the log-expression scale), which motivates our choice to evaluate estimated expression rather than estimated $\beta_1$.

\begin{figure}[!htb]
	\centering
	\mbox{
		\centering
		\includegraphics[width=.45\linewidth]{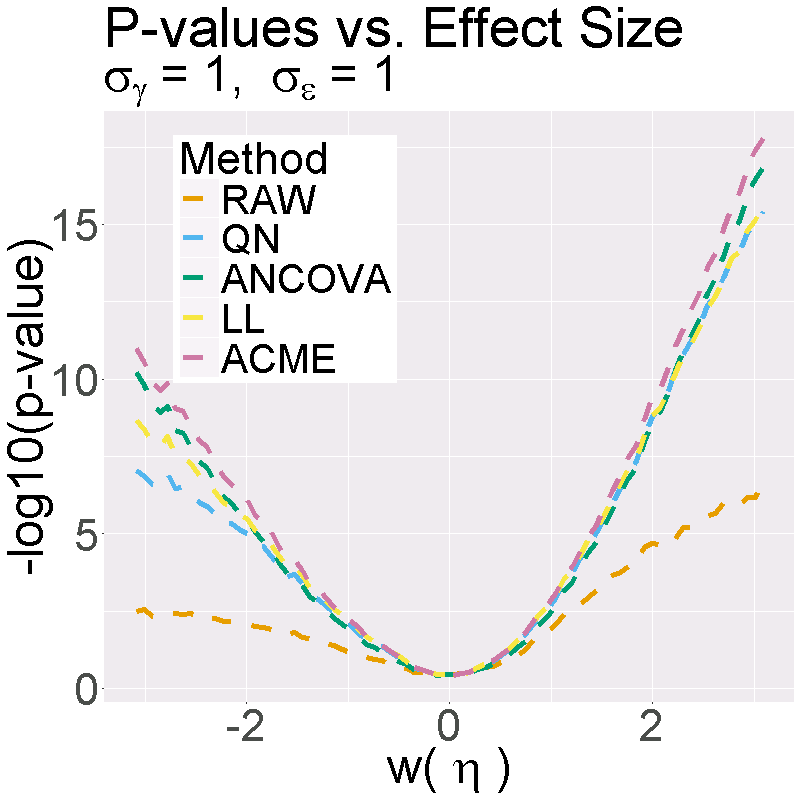}
	}
	\mbox{
		\includegraphics[width=.45\linewidth]{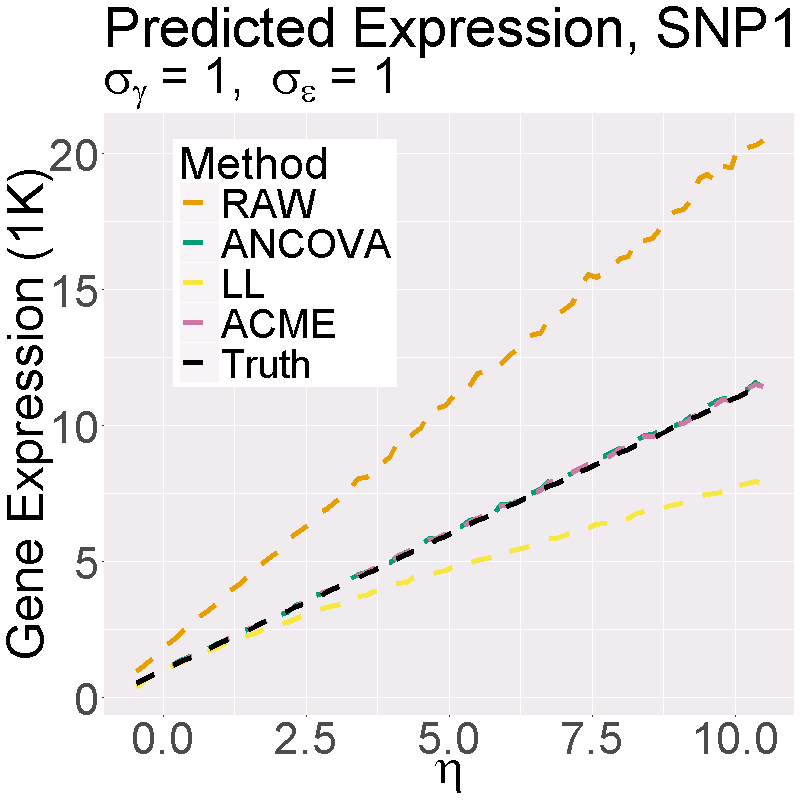}
		\includegraphics[width=.45\linewidth]{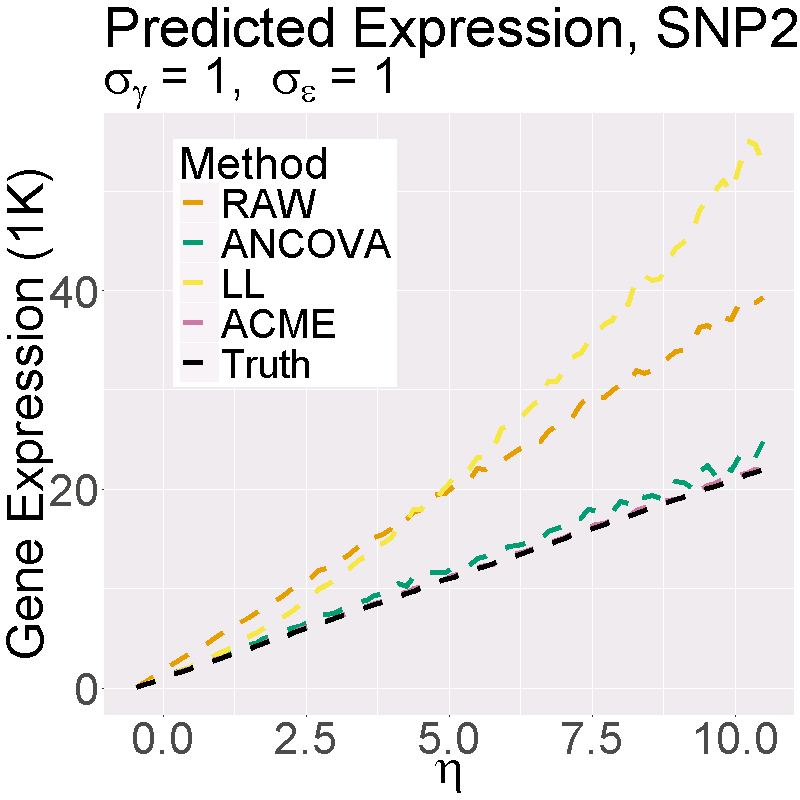}
	}
	\caption{Results of large-scale simulation experiment. Middle: $-\log{10}$ $F$-test p-values as a function of $\eta$. Left and right: predicted raw expression with one and two reference alleles, respectively.}
	\label{fig:power-sims}
\end{figure}

In Figure \ref{fig:power-sims}, we display results from the above simulation framework with $\sigma_\eps = \sigma_\gamma = 1$. Note that, for the top plot, the $x$-axis is transformed according to $w(\eta) := \log(1 + 2\eta)$, a scale which better displays the results when $\eta\in(-0.5, 0)$. We replicated the simulation framework with other choices of $\sigma_\gamma$ both above and below $\sigma_\eps$. The results of these replications are shown in Web Figure \ref{fig:power-sims2}. Our results show that the ACME model achieves the most power and estimation accuracy among the alternative methods. Hence, if the ACME model is the best representation of underlying eQTL biology in terms of allele count (as the analyses in Section \ref{diagnostics} suggest), use of the existing methods reduces accuracy and sensitivity. Additionally, in Web Figure \ref{fig:power-sims2}, we show that when $\sigma_\gamma$ is increased (i.e., the covariates have more effect), the accuracy and sensitivity of the competing methods is even further reduced.

\textbf{Remark}: The power of log-ANCOVA is quite close to that of ACME. This is expected, as log-ANCOVA contains ACME, the true simulated model. Thus, in this simulation study, the log-ANCOVA results act mainly as a reference; the larger model's practical and conceptual shortcomings compared to ACME were discussed in Section \ref{Model}. The focus of this study is mainly on single-parameter models and the consequences of misspecification.

\subsection{Computation times}\label{Power:computation}
To assess computation speed, we timed various methods on every simulation instance for Figure \ref{fig:power-sims}. There were 10,000 unique values of $\eta$, and therefore 1 million simulation instances. On each instance, we recorded the computation time for: (1) the LL model with least-squares estimation; (2) the ACME model with maximum likelihood using the BFGS method implemented in $\mathtt{optim}$  from $\mathtt{R}$; (3) the ACME model with the custom fitting algorithm derived in Web Appendix \ref{Fitting}.
The timing of (1) will be of the same order as any other procedure based on least-squares (RAW, QN, and ANCOVA). The timing of procedure (2) is provided as a benchmark for procedure (3). All computations were performed on an Intel Xeon E5-2640 (2.50 GHz), using the $\mathtt{R}$, and timed with the $\mathtt{microbenchmark}$ package. 

The mean and standard deviation of computation times for the procedures, over the 1 million simulation instances, were as follows: least-squares at 0.129ms (0.233), BFGS at 2.687ms (1.270), custom ACME at 0.470ms (0.285). So, the efficiency of the custom ACME algorithm is quite comparable to that of least-squares estimating equations, and outstrips stock optimization methods. The complete package implementing our algorithm, including wrappers that employ parallelization to process all cis-eQTL results from massive-scale eQTL data, is available in the $\mathtt{ACMEeqtl}$ package on the CRAN respository.




\section{Large-scale real data analysis}\label{RealData}

\begin{figure}[!htb]
\centering
\mbox{
	\includegraphics[width=.45\linewidth]{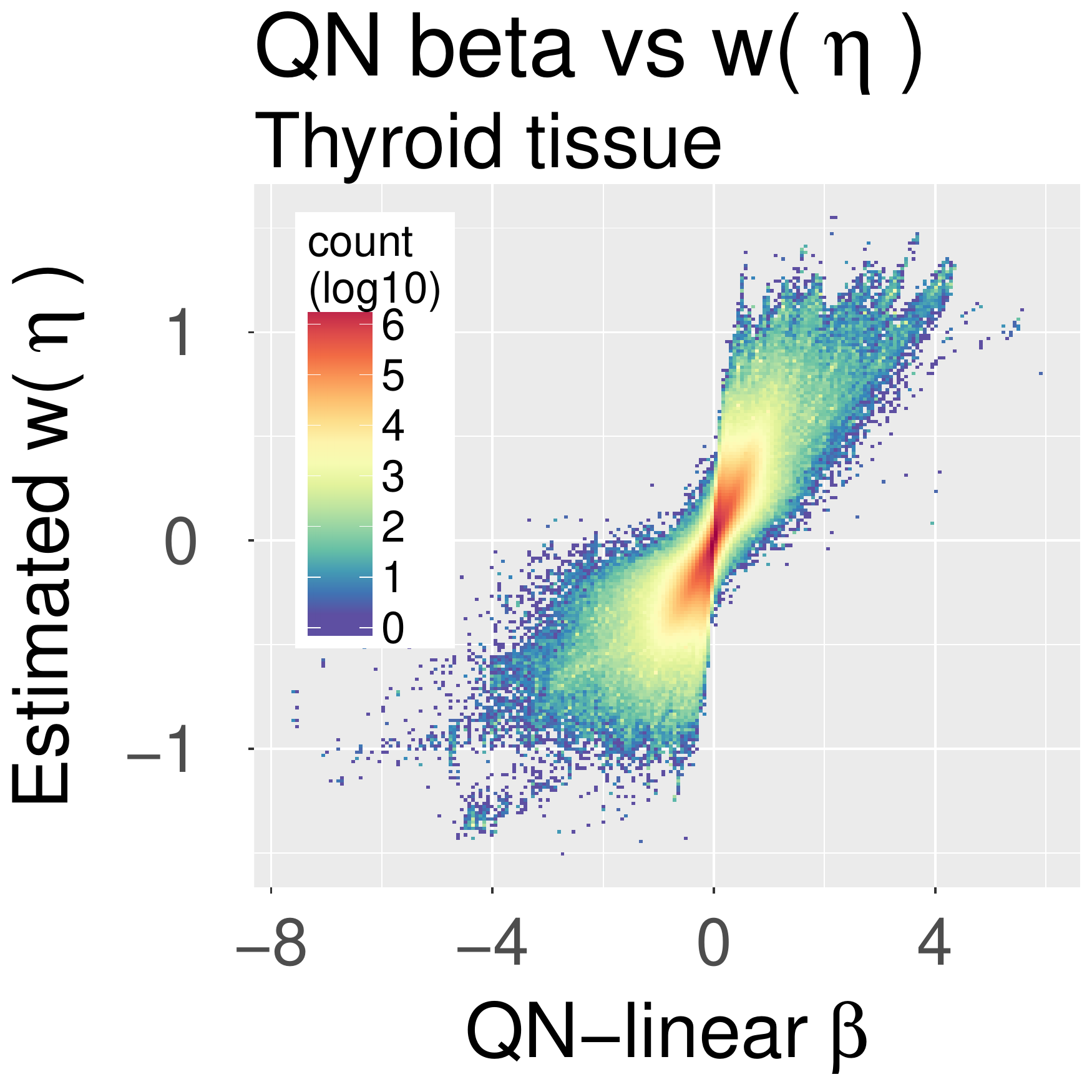}
	\includegraphics[width=.45\linewidth]{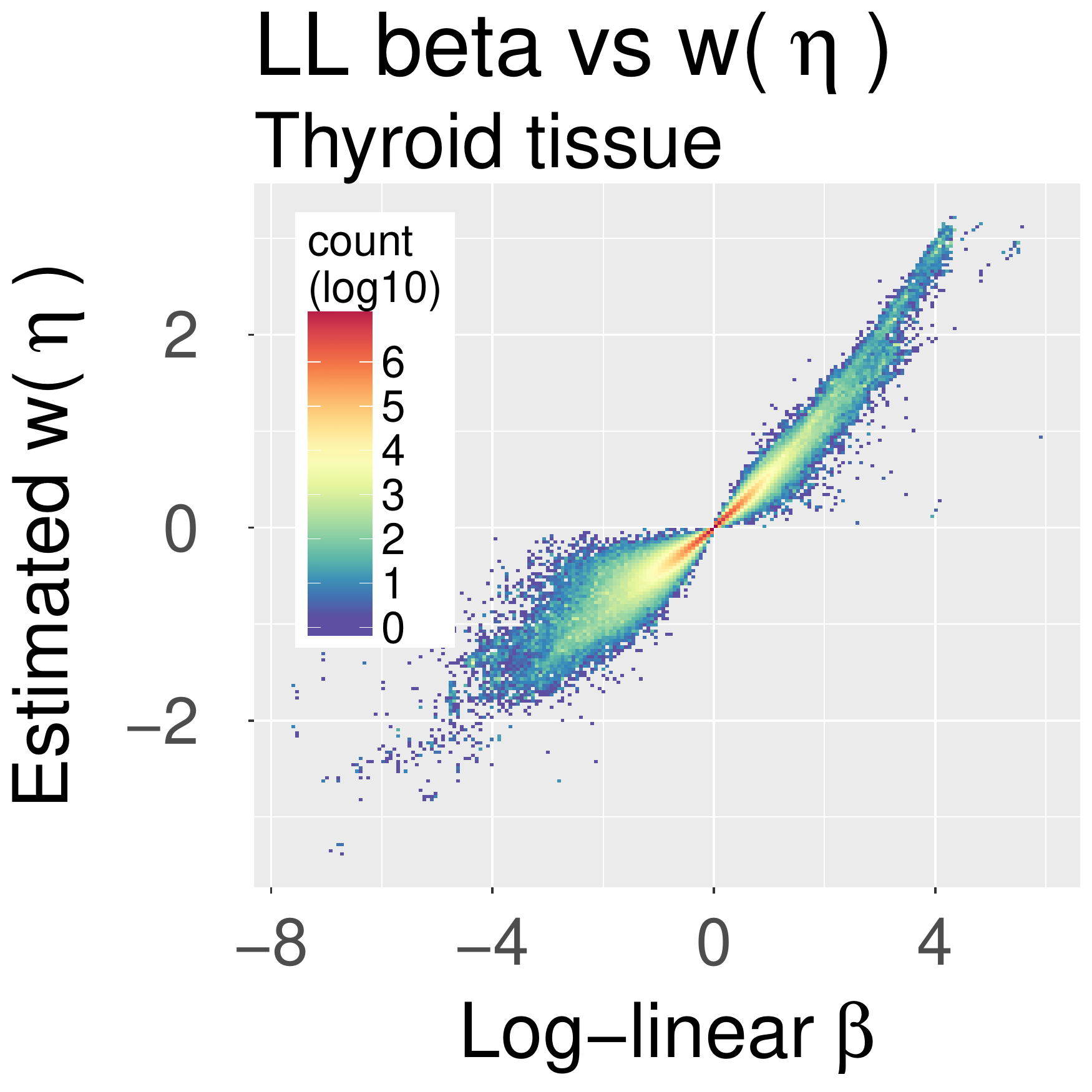}
}
\vskip0.2cm
\mbox{
	\includegraphics[width=.45\linewidth]{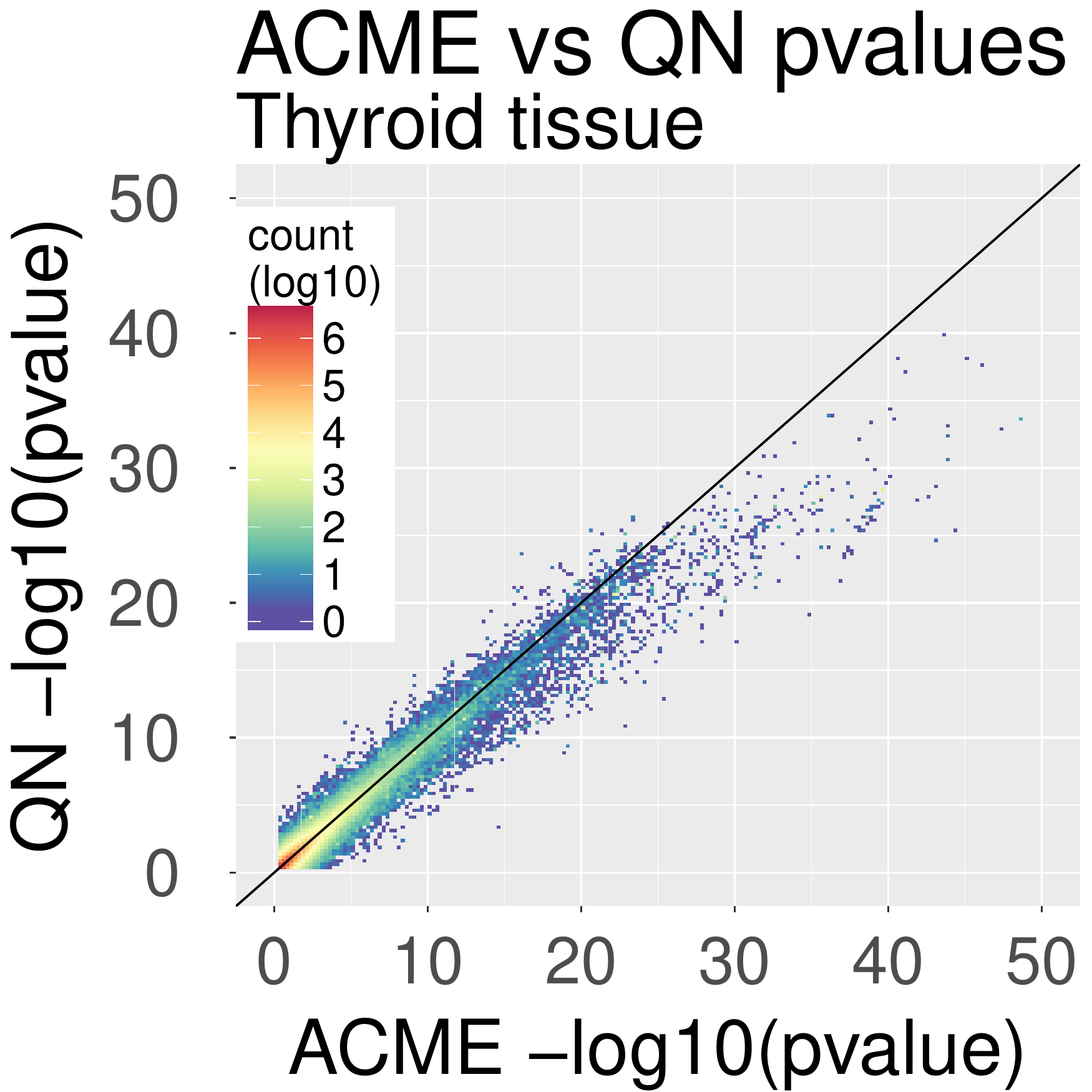}
  \includegraphics[width=.45\linewidth]{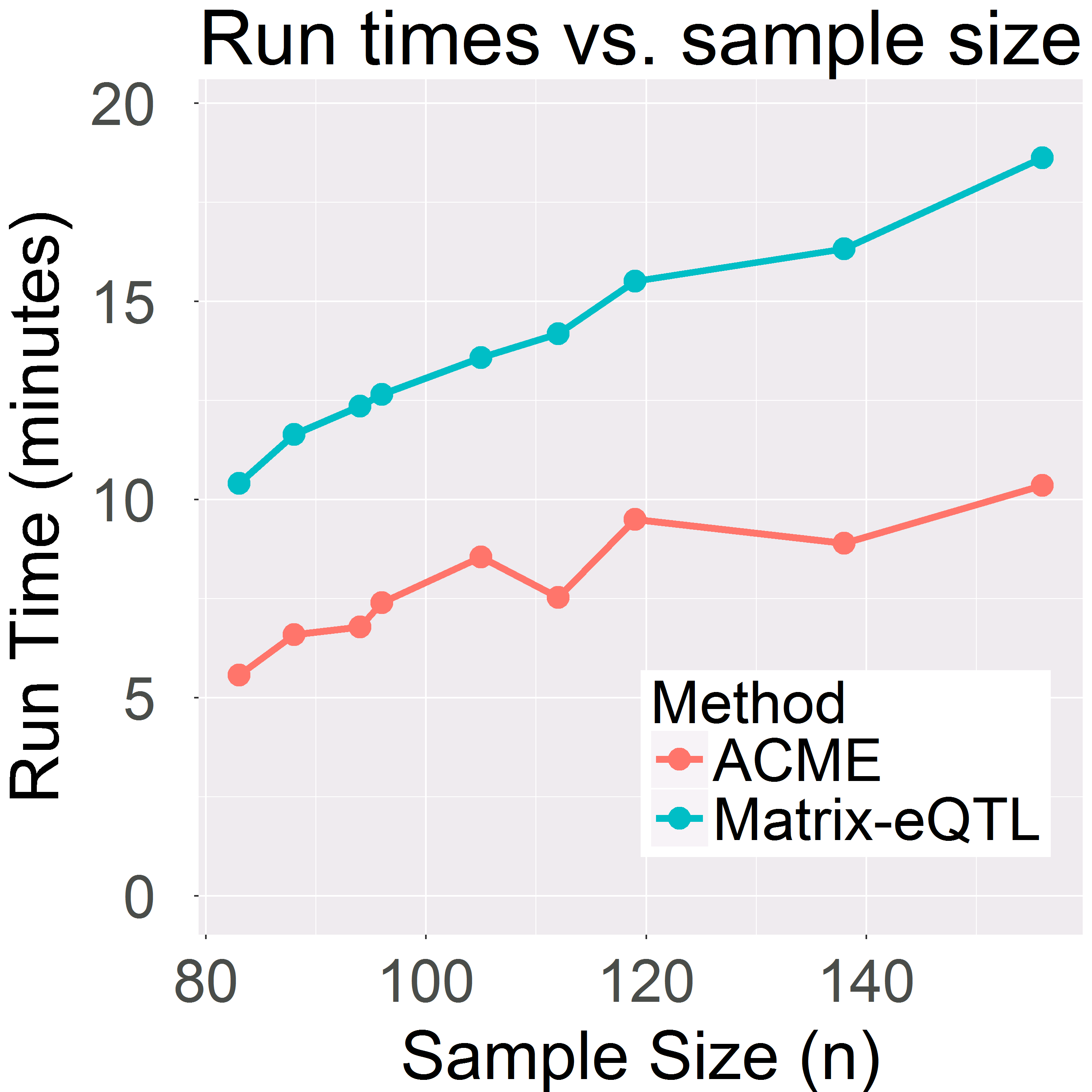}
}
\caption{\label{fig:full3} Results of genome-wide cis-eQTL ACME effect size estimations on Thyroid tissue ($n = 105$) from GTEx data. Top row: comparisons of ACME effect sizes (transformed with $w$ as introduced in Section \ref{Power}) with both QN and LL effect sizes. Bottom-left: QN vs ACME regression p-values. Bottom-right: Full-tissue procedure times of the Matrix-EQTL and ACME fitting softwares.}
\end{figure}

This section contains further comparisons between different effect-size models. Note that the most important real-data comparisons between ACME and standard models are the goodness-of-fit tests shown in Section \ref{diagnostics}. Those tests showed that ACME best fits cis-eQTL data, compared to other allele-count models. Moreover, ACME estimates correspond to a coherent biological interpretation of cis-eQTL action. In light of this, it can be asked whether effect-size \emph{rankings} from other methods, at least, correspond to those from ACME. Effect size estimates for all cis-pairs were computed from Thyroid tissue data using the QN-linear, LL, and ACME models. The top plots in Figure \ref{fig:full3} show that effect-size ordering given to the strongest eQTLs by QN and LL differ markedly from ACME. Also in Figure \ref{fig:full3} is a plot of ACME versus QN-linear regression $p$-values, shown mainly to provide a fuller comparison to \cite{ardlie2015genotype} and other studies which rely on the QN-linear model. It is clear that, while QN p-values are associated with ACME's (as one would hope), they are by no means identical in rank, and quite different in magnitude for the strongest eQTLs. Furthermore, as shown in the top row of plots, many QN effect sizes differed in sign from ACME's. All these comparisons held across tissues, as shown in Web Figure \ref{fig:full-extra}. Beyond method comparisons, we also display summaries of ACME effect sizes in Web Figure \ref{fig:full-extra2}.
Computation times for real-data analyses were recorded for both the ACME and QN model fitting procedures, using the same PC as mentioned in Section \ref{Power:computation}. The $\mathtt{R}$ packages $\mathtt{MatrixEQTL}$ and $\mathtt{ACMEeqtl}$ provide full-tissue cis-eQTL procedures for the QN and ACME models (respectively). We timed the run of each procedure on each of the nine tissues. The results are shown in the bottom-right of Figure \ref{fig:full3}. We note that the ACME software benefits from parallelization that the Matrix-EQTL software currently does not employ. This explains why the ACME full-tissue procedure is faster than Matrix-EQTL, even though the former is based on an iterative optimization procedure.




\section{Discussion}\label{Discussion}

We have proposed ACME, a new model for the effect-size of cis-eQTLs. ACME follows a simple additive model for cis-eQTL action, and is supported by careful analysis of real data. 
In particular, we show via goodness-of-fit tests that while the error distribution of gene expression is best modeled on the log-scale, cis-eQTL additive allelic effects occur on the \emph{original} scale, contrary to assumptions implied by standard transformations.
Beyond simply harmonizing some biological considerations regarding allele-specific expression, this analysis shows that a single parameter can be used in most instances to catalog the effect size of a cis-eQTL, and that dominance in cis-eQTLs appears to be rare.
Simulations in Sections \ref{pvalues} and \ref{Power} showed the robustness and the superior power of the ACME model. Real-data analyses in Section \ref{RealData} suggested that, for eQTL ranking purposes, estimates and p-values from standard models (QN and LL) are not adequate stand-ins for those from ACME. Furthermore, as seen in the goodness-of-fit tests, use of the QN-linear or LL models can create false evidence of dominance. Finally, we showed ACME estimation on all cis-pairs is computationally feasible, and have provided open-source software.

We believe the ACME model places cis-eQTL effect size analysis on a solid statistical foundation, and can be readily implemented in current eQTL studies. The results may be useful for investigations in which interpretable eQTL effect sizes and reliable rankings are relevant, such as examining enrichment and overlap with genome-wide association studies \citep{zhu2016integration}. Furthermore, our analytic standard errors for ACME effect sizes allows ACME estimates to be used directly in methods for downstream analysis of multi-tissue eQTL variation \citep{flutre2013statistical, li2013empirical}.

Though we did not explore the application of the ACME model to trans-eQTL analysis, the ACME model can in principle be applied to trans-eQTLs. However, for trans-eQTLs, dominance effects may be more plausible, while current sample sizes may be inadequate to fully investigate such effects. Thus we consider the use of the ACME model for trans-eQTLs to be exploratory. Another possible extension is a multi-SNP ACME model. However, it is not obvious that allelic additivity  should hold in a multi-SNP model. 
Nonetheless, to facilitate analysis, we have implemented a step-wise fitting algorithm for a multi-SNP ACME model using additivity across loci, with code included in our software package. These estimates can be used to consider preliminary ACME models that are additive in genotypes across SNPs, pending more rigorous development of a multi-SNP approach.

\backmatter


\section*{Supplementary Materials}

Web Appendices and Web Figures referenced in Sections \ref{existing}, \ref{notation}, \ref{Model}, \ref{log-models}, \ref{ACME}, \ref{diagnostics}, \ref{F-test-sims}, \ref{importance-sampling}, \ref{Power}, \ref{Power:computation}, and \ref{RealData} are available with
this paper at the Biometrics website on Wiley Online Library. Code to fit the ACME model comes with the $\mathtt{ACMEeqtl}$ package, available on the CRAN repository.\vspace*{-8pt}

\section*{Acknowledgements}

GTEx data were from dbGaP phs000424.v3.p1 (http://www.gtexportal.org),  supported by the NIH Common Fund (commonfund.nih.gov/GTEx). Supported in part by NIH R01MH101819-01, HG007840, NSF DMS1310002, and EPA RD83574701. The authors acknowledge many helpful conversations and phone calls with members of the GTEx Consortium.\vspace*{-8pt}



%

 \bibliographystyle{biom} 
 \bibliography{Refs}



















\end{document}